\documentclass[twocolumn]{aastex62}

\graphicspath{{./}{figures/}}

\shorttitle{MWL DA~495}
\shortauthors{Coerver, Wilcox, and Zhang, et al.}

\usepackage{gensymb}
\usepackage{textcomp}
\usepackage{float}
\usepackage[caption=false]{subfig}
\usepackage{verbatim}
\usepackage{amsmath}
\newcommand{\textapprox}{\raisebox{0.5ex}{\texttildelow}}
\newcommand{\nustar}{\textit{NuSTAR}}
\newcommand{\chandra}{\textit{Chandra}}
\newcommand{\xmm}{\textit{XMM-Newton}}

\begin{document}

\title{Multiwavelength Investigation of Pulsar Wind Nebula DA~495 with \textit{HAWC}, \textit{VERITAS}, and \textit{NuSTAR}}

\correspondingauthor{Patrick Wilcox}
\email{patrick-wilcox@uiowa.edu}

\author{Coerver, A.}
\affiliation{Department of Physics and Astronomy, Barnard College, Columbia University, NY 10027, USA}
\author{Wilcox, P.}
\affiliation{Department of Physics and Astronomy, University of Iowa, Van Allen Hall, Iowa City, IA 52242, USA}
\author{Zhang, H.}
\affiliation{Department of Physics and Astronomy, Purdue University, West Lafayette, IN 47907, USA}
\author{Dingus, B.L.}
\affiliation{Los Alamos National Laboratory, Los Alamos, NM, USA}
\author{Gotthelf, E.V.}
\affiliation{Columbia Astrophysics Laboratory, Columbia University, New York, NY 10027, USA}
\author{Hailey, C.J.}
\affiliation{Columbia Astrophysics Laboratory, Columbia University, New York, NY 10027, USA}
\author{Humensky, T.B.}
\affiliation{Physics Department, Columbia University, New York, NY 10027, USA}
\author[0000-0002-3638-0637]{Kaaret, P.}
\affiliation{Department of Physics and Astronomy, University of Iowa, Van Allen Hall, Iowa City, IA 52242, USA}
\author{Li, H.}
\affiliation{Los Alamos National Laboratory, Los Alamos, NM, USA}
\author[0000-0002-9709-5389]{Mori, K.}
\affiliation{Columbia Astrophysics Laboratory, Columbia University, New York, NY 10027, USA}
\author[0000-0002-3223-0754]{Mukherjee, R.}
\affiliation{Department of Physics and Astronomy, Barnard College, Columbia University, NY 10027, USA}
\author[0000-0002-4282-736X]{Park, N.}
\affiliation{WIPAC and Department of Physics, University of Wisconsin-Madison, Madison, WI, USA}
\author{Zhou, H.}
\affiliation{Los Alamos National Laboratory, Los Alamos, NM, USA}



\begin{abstract}
Pulsar Wind Nebula (PWN) DA~495 (G65.7+1.2) was detected in TeV gamma-rays by the High Altitude Water Cherenkov Observatory (HAWC) in 2017 (2HWC~J1953+294). Follow-up observations by the Very Energetic Radiation Imaging Telescope Array System (VERITAS) confirmed the association between 2HWC~J1953+294 and DA~495 and found the TeV emission to be spatially coincident with the radio emission first reported in 1968. The detection of TeV gamma-rays from DA~495, along with past X-ray detection up to 10 keV, prompted high energy X-ray observations as part of the \nustar\ Galactic Legacy Survey. We present the results of these \nustar\ observations, combined with archival \chandra\ and \xmm\ observations, and confirm the previous X-ray photon index of $\Gamma_{2-20 \rm\ keV} = 2.0 \pm 0.1$. We find no spectral cutoff up to 20 keV. With the spectral information for DA~495 extended to TeV gamma-rays, we were able to perform analytical modeling to test leptonic and hadronic emission scenarios. The leptonic models can explain the broadband emission, but also imply a diffuse X-ray nebula of similar extent to the radio and TeV nebulae, which cannot be confirmed by our observations. The hadronic models can simultaneously explain the spectrum and the spatial extent in all wavelengths; however, we need a very high magnetic field strength pervading the radio and TeV nebulae and a surprisingly high particle kinetic energy. These requirements deepen the mystery of the physical nature of DA~495. Future observations in radio to infrared bands and spatially resolved $\gamma$-rays can further constrain the physical conditions and radiation mechanisms in DA~495.

\end{abstract}



\section{Introduction} \label{sec:intro}


When a massive star ($>8\ M_\sun$) explodes in a core collapse supernova, much of its rotational energy remains with the newly created neutron star (pulsar). This pulsar can produce an expanding bubble of highly relativistic wind, called the pulsar wind nebula (PWN), and forms a termination shock. The observation of synchrotron and inverse Compton emission within the PWN suggest the acceleration of nonthermal particles. The pulsar may continue to feed the PWN with its rotational energy and expand the boundary of the nebula. For a review of PWNe see \citet{2006ARA&A..44...17G}.

The most well studied (and thought to be representative) PWNe are the young Crab and older Vela-X Nebulae with ages of about 1,000 years and 20,000 years, respectively \citep{2008ARA&A..46..127H,2015SSRv..191..391K}. Vela-X is notably more evolved with its very extended filamentary structure within the PWN and apparent expanding supernova remnant (SNR) shell seen to interact with the local interstellar medium (ISM). This shell, evidence of an advancing shock formed by the progenitor supernova, is thus far absent from observations of the Crab Nebula \citep{2015ApJ...806..153Y}. Both the Crab and Vela PWNe have well studied pulsars (PSR B0531+21 and PSR B0833-45, respectively) which allow the broadband emission to be modeled and measured within the context of the pulsar-nebula system \citep{1988ApJ...332..199S,2016MNRAS.462.2762B}. Studies of the Crab have indicated Lorentz factors of electrons reaching $\gamma \approx 10^6$ \citep{2017MNRAS.471.4856F}, emphasizing the importance of PWNe as astrophysical laboratories for relativistic processes. This is particularly important when considering the contribution of nearby PWNe to the cosmic ray positron population \citep{2018PhRvD..98f2004A,2009APh....32..140G}. Much of our knowledge about PWN evolution comes from comparing these two objects, however more detail about evolutionary processes is clearly necessary, as environmental and progenitor characteristics play key roles in the development of PWNe. Since not all PWN have age and pulsar characteristics available or as much environmental context as the Vela-X nebula, we must rely on spectral modeling and analysis of energy-dependent morphology to explain the conditions of these astrophysical laboratories and try to place them within an evolutionary context of other well-studied PWN. In this study we present results from multiwavelength modeling of PWN DA~495 and evaluate its energy dependent morphology using new observations from \nustar. It is thought that DA~495 is in an evolutionary state somewhere between the Crab and Vela-X PWNe \citep{2008ApJ...687..516K}, making it a good candidate for investigation of PWNe life cycles. In section \ref{sec:background} we discuss the observational history of DA~495, including: radio observations, X-ray observations, and recent very high energy gamma-ray observations not used in past modeling studies. Section \ref{sec:observations} details the new \nustar\ observations that are analyzed and reported on in Section \ref{sec:results}. Section \ref{sec:model} describes models and derived parameters used to evaluate the nebula, and Section \ref{sec:discussion} discusses implications of the modeling results.

\section{Background/Previous Observations} \label{sec:background}

\subsection{Discovery and identification}
DA~495 (G65.7+1.2) was discovered by the Dominion Astrophysical (DA) survey \citep{1968AJ.....73..135G}. It was identified as a point source in the DA survey due to the coarse angular resolution but there was insufficient signal to determine any spectral information at the time. Follow up as part of a supernova remnant search, performed with the National Radio Astronomy Observatory (NRAO) and  the Vermilion River observatory, found that DA~495 had an extended structure and a non-thermal spectrum suggestive of a Crab-like SNR \citep[and erratum]{1973A&A....26..237W}. Further confirmation of the center-filled SNR hypothesis came with additional Dominion Radio Astrophysical Observatory (DRAO) observations \citep{1983AJ.....88.1810L}. We now know these center-filled SNRs to be PWNe, and DA~495 has been studied as such since then, however an associated pulsation has not been found. More recent results from observations in radio and other wavelengths are detailed in the next sections.

\subsection{Radio}
\begin{figure}%
    \centering
    \includegraphics[width=0.9\columnwidth]{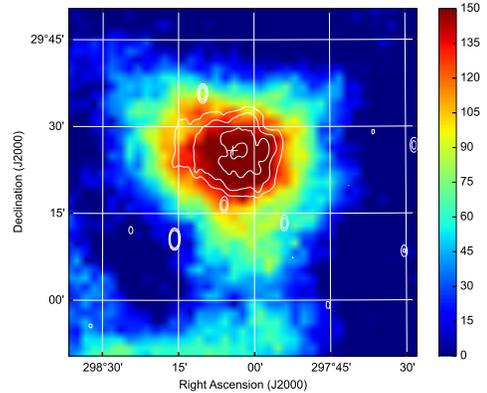}
    \caption{The color image is VERITAS excess $>200$ GeV from \citet{2018ApJ...866...24A} and is in units of counts$/ 0.09 \rm\ deg^2$. Contours are $\rm T_b=(8,9,11)\ K$ from 1420 Mhz Canadian Galactic Plane Survey \citep{2003AJ....125.3145T}. The cross marks the PWN center in X-rays from this work. Note that the radio contours of the nebula from outer to inner are increasing then decreasing: 8,9,11,9 K, indicating the radio hole in the center. The VERITAS excess map is integrated with $\theta < 0.3\deg$, so the 2$\arcmin$ radio hole, if present, would not be apparent in the gamma-ray map. }%
    \label{fig:historyskymap}%
\end{figure}

The most recent radio analysis of DA~495 is reported in \citet{2008ApJ...687..516K}, hereafter referred to as K08. The maps generated by K08 from observations done with the Canadian Galactic Plane Survey in 408 MHz and 1420 MHz \citep{2003AJ....125.3145T} and the Effelsberg Radio Telescope in 4850 MHz and 10550 MHz show an approximately circular diffuse source of about 25$\arcmin$ in diameter. That size is fairly consistent across the radio spectrum reported in K08 and is consistent with earlier observations, but the nebula is not present in IRAS $60 \rm\ \mu m$ observations. Fractional polarization is about 25\% at higher frequencies (2695 MHz, 4850 MHz and 10550 MHz) and is ordered in a way that indicates a central dipole magnetic field with a superimposed toroidal component. The magnetic field of the nebula, measured from a synchrotron cooling break at 1.3 Ghz, was determined to be extremely high at $B=1.3\ \rm mG$. Using HI absorption and kinematic measurements relating to galactic rotation, K08 also estimated a distance to the source of $d=1.0 \pm 0.4\ \rm kpc$. We adopt the radio fluxes reported in K08, which have compact sources and the foreground H II region removed for use with our combied spectrum described in Section \ref{sec:model}. A map of radio emission at 1.4 GHz from the Canadian Galactic Plane Survey (CGPS) is shown as contours in Figure \ref{fig:historyskymap}.

\subsection{X-ray}
DA~495 was first detected in the X-ray band in March 2004 using archival \textit{ROSAT} and \textit{ASCA} data \citep{Arzoumanian2004}. \textit{ROSAT} source 1WGA J1952.2+2925, a faint, compact X-ray source, was assumed to be associated with the surrounding radio nebula. The X-ray flux was found to be non-variable, non-thermal, and extended, leading to its identification as emission from a wind nebula. DA~495 was followed up with \chandra\ in 2007 \citep{2008ApJ...687..505A}, which was able to resolve a central point source inside the \textapprox $40"$ diameter (0.2 parsec at a distance of 1 kpc) X-ray nebula. \chandra\ spectral analysis of the extended emission resulted in a photon index of $\Gamma = 1.6 \pm 0.3$, allowing for confirmation of the extended region as a wind nebula. The central point source was found to have a purely thermal spectrum, a result confirmed by \cite{Karpova} (hereafter K15). K15 jointly fit the archival \xmm\ and \chandra\ data with an absorbed powerlaw plus blackbody model and an absorbed powerlaw plus neutron star atmosphere model (NSMAX) \citep{Mori2007,Ho2008} to characterize the point source emission. The blackbody fit, likely modeling emission from a polar cap, resulted in a temperature of $T \approx 0.22$ keV, a radius of $R \approx 0.6$ km, and a neutral hydrogen column density of $N_H \approx 2.6 \times 10^{21}$ cm\textsuperscript{-2}. The NSMAX fit, modeling emission from the entire neutron star surface, resulted in a temperature of $T \approx 0.08$ keV, a radius of $R \approx 10$ km, and a neutral hydrogen column density of $N_H \approx 3.5 \times 10^{21}$ cm\textsuperscript{-2}. Neither model was ruled out. No pulsations were detected in the \chandra\ data, but the pure thermal spectrum of the point source implies its likely identification as the neutron star powering the wind nebula. K15 further constrained the pulsation non-detection through timing analysis of the \xmm\ data, setting the upper limit for a pulsed fraction of 40 percent in a range of $\geq 12.5$ ms. K15 estimated the distance to the putative neutron star to be \textapprox2.4 - 3.3 kpc (depending on which spectral models were applied to the point source spectrum) using the $N_H-D$ relation. 

\subsection{Gamma ray}
The first TeV gamma-ray detection in the region of DA~495 was reported by the High Altitude Water Cherenkov Telescope (HAWC), which detected  2HWC J1953+294 as a point source within 0.2\degr\ of the radio center reported by K08 \citep{2017ApJ...843...40A}. Given the 25' diameter radio size of DA~495 and HAWC's localization uncertainty of 0.1\degr\, it is certainly plausible that DA~495 is associated with 2HWC J1953+294 by position alone. Additional evidence that 2HWC J1953+294 is associated with DA~495 is the photon index at 7 TeV of $\Gamma = 2.78 \pm\ 0.15$, which is consistent with other PWNe in this energy range (such as the Crab). The TeV gamma-ray detection of DA~495 was confirmed by follow-up observations with the Very Energetic Radiation Imaging Telescope Array System (VERITAS), which detected VER J1952+293 \citep{2018ApJ...866...24A}. VER J1952+293 is centered within 0.05\degr\ of the radio center and has a extension defined by a 2D Gaussian with $\sigma = 0.14\degr\ \pm\ 0.02\degr$, which is consistent with the size of the radio nebula. The extent and position of the emission detected by VERITAS provides compelling evidence that the TeV and radio nebula are associated. The photon index measured by VERITAS at 1 TeV, $\Gamma = 2.65 \pm\ 0.49$, also agrees well with the TeV PWN interpretation.

There is significant disagreement between the VERITAS and HAWC flux measurements of their respective sources. At 1 TeV, HAWC has a flux that is about 7$\times$ higher than that of VERITAS \citep{2018ApJ...866...24A} and is probably attributed to a nearby extended source and/or diffuse emission that contaminate the flux reported for 2HWC J1953+294. For this study, we adopt the VERITAS flux points and accept the HAWC measurement as an upper limit.



\section{Observations} \label{sec:observations}
\subsection{X-ray Observations}

Observations of DA~495 were performed by \nustar\ on 2017 June 9 (ObsID 30362003002) for a total exposure of 60 ks. \nustar\ pointed at the X-ray centroid of the PWN at R.A.(J2000) = 19\textsuperscript{h} 52\textsuperscript{m} 17.04\textsuperscript{s}, decl.(J2000) = 29\degree\ 25' 52.5" \citep{2008ApJ...687..505A}. The \nustar\ data was processed with \texttt{nupipeline 0.4.6}. DA~495 was observed by \chandra/ACIS-I on 2002 December 9 (ObsID 3900) for a total of 25 ks. Observations also took place with \xmm\ EPIC/MOS (Full Frame Mode, medium filter setting) on 2007 April 21 for a total of about 50 ks. \xmm\ data was processed using \texttt{XMM SAS 1.2}. 

The \nustar, \chandra, and \xmm\ datasets were used for both image analysis and spectroscopy. All three datasets underwent spectral extraction and joint fitting in \texttt{XSPEC} (v12.9.0), providing a more complete picture of the DA~495 X-ray emission.

\section{Results} \label{sec:results}
\subsection{Data Analysis}
\subsubsection{\nustar\ Data Reduction}

The \nustar\ data was processed and analyzed using the \texttt{HEASOFT V6.21} software package, including \texttt{NUSTARDAS 06December16 V1.7.1}, with \nustar\ Calibration Database (CALDB) files from 2017 June 14. 

\nustar\ background-subtracted images were obtained using the \texttt{nuskybgd} software \citep{Wik2014}. \texttt{Nuskybgd} maps and generates images for the entire background, allowing for specific background spectrum extraction given a source region. It takes into account stray light leaking through the aperture stop, focused cosmic X-ray background, instrumental background, and soft environmental neutrons from cosmic rays. When modeling the background with \texttt{nuskybgd}, three source-free regions were selected from each module, each region on a different detector chip. Although stray light contamination mainly manifested below 3 keV, the affected region, the top third of module A, was avoided when choosing background regions for modeling. Near-source background light curve analysis revealed no periods of high background or flaring.

\begin{figure}%
    \centering
    \subfloat[]{\includegraphics[width=0.45\textwidth]{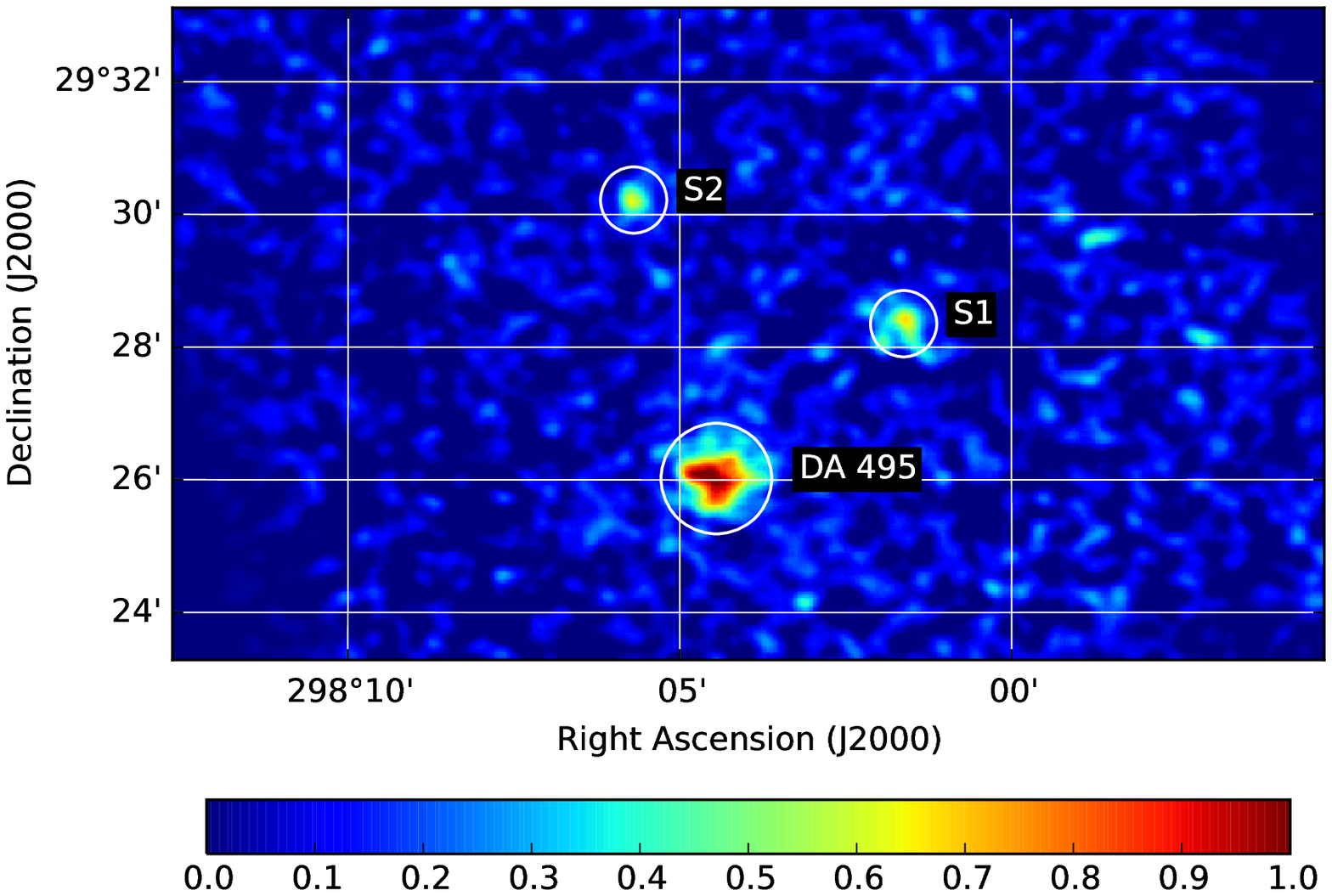}}%
    \qquad
    \subfloat[]{{\includegraphics[width=0.45\textwidth]{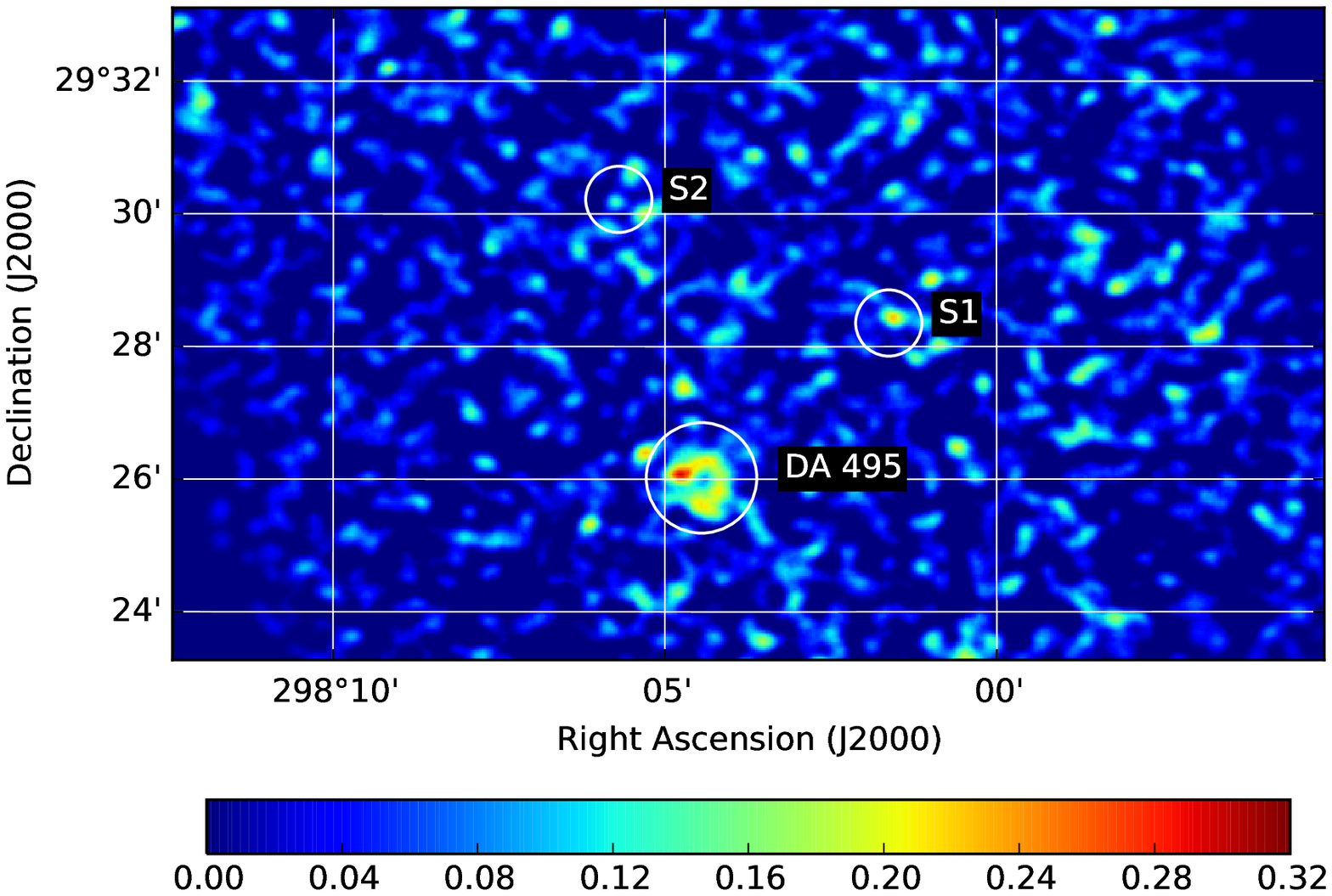} }}%
    \caption{(a) \nustar\ modules A and B from 3 to 20 keV, summed and background subtracted. (b) \nustar\ modules A and B from 10 to 20 keV, summed and background subtracted. The images were smoothed with a Gaussian kernel to $5\sigma$ significance.}%
    \label{fig:nustar}%
\end{figure}

After background subtraction three X-ray sources were visible in the \nustar\ field of view (Fig. \ref{fig:nustar}a). \texttt{CIAO wavdetect} detected the two point sources, S1 and S2, in the \nustar\ data from 3 to 10 keV, but did not detect them in the 10 to 20 keV band. DA~495 appears up to 20 keV. The archival \chandra\ data revealed counterpart point sources to S1 and S2: CXO J195205.6+292808 (S1) and CXO J195222.6+293005 (S2). These nonvariable low-energy point sources are extremely unlikely counterparts to the TeV gamma-ray emission and are excluded from further analysis.

\subsection{X-ray Spectral Analysis}

We extracted spectra from both the \nustar\ data and archival \chandra\ and \xmm\ data and jointly fit the low- and high-energy spectra.

\subsubsection{\nustar}

For \nustar\ spectral extraction with \texttt{nuproducts 0.3.0}, a region of r \textless\ 50" was used, centered at the PWN centroid in the full band. The extraction region was increased beyond the X-ray extent of  r \textapprox\ 20", as the larger \nustar\ PSF can cause counts to spill beyond the \chandra-measured source extent. A 50" region, determined to be optimal through image analysis and trial-and-error spectral extraction, was large enough to include all or nearly all source counts while preserving a high signal-to-noise ratio. Using \texttt{nuproducts} we generated the \nustar\ response matrix (RMF) and effective area (ARF) files for an extended source. Extracted spectra from module A and B were found to have consistent fluxes and were combined using \texttt{addspec} from \texttt{FTOOLS 6.9}. The spectrum was subsequently binned to 2$\sigma$ significance over background counts in each bin. 

\nustar\ background spectra were generated by jointly modeling module A and B with \texttt{nuskybgd}. Background spectrum generation with \texttt{nuproducts} was also attempted. \texttt{Nuproducts} uses only one source-free rectangular background region file per module, which we selected to be on the same detector chip as the source region. Although fitting the source spectra with both \texttt{nuskybgd} and \texttt{nuproducts} background spectra yielded largely consistent results, \texttt{nuskybgd} produced a slightly better match between the A and B spectra and was used for final fitting. Like the source spectra, \texttt{addspec} was used to combine the module A and B background spectra generated with \texttt{nuskybgd}.

\subsubsection{\chandra}

\chandra\ spectral extraction was performed using \texttt{CIAO 4.10} procedures for extended emission. An extraction radius of r \textless\ 20", the \chandra-measured source extent \citep{Karpova}, was used. A background spectrum was extracted from a circular, 70" radius, point-source-free region on the same detector chip as DA~495. The extracted source spectrum was variably binned to 2$\sigma$ significance.

\subsubsection{\xmm}

\xmm\ spectral extraction was performed using \texttt{XMM SAS 17.0.0} procedures. A region of r \textless\ 40" was used for spectral extraction. As \xmm\ has a larger PSF than \chandra, a larger region was required to capture all source counts. The background spectrum was extracted from a 70" circle covering a nearby region that was determined to be point-source-free by \texttt{XMM SAS edetect\textunderscore chain}. The extracted source spectrum was variably binned to 2$\sigma$ significance. 

\subsubsection{Fitting Results}

\begin{figure}%
\centering
    \includegraphics[width=0.6\columnwidth,angle=-90]{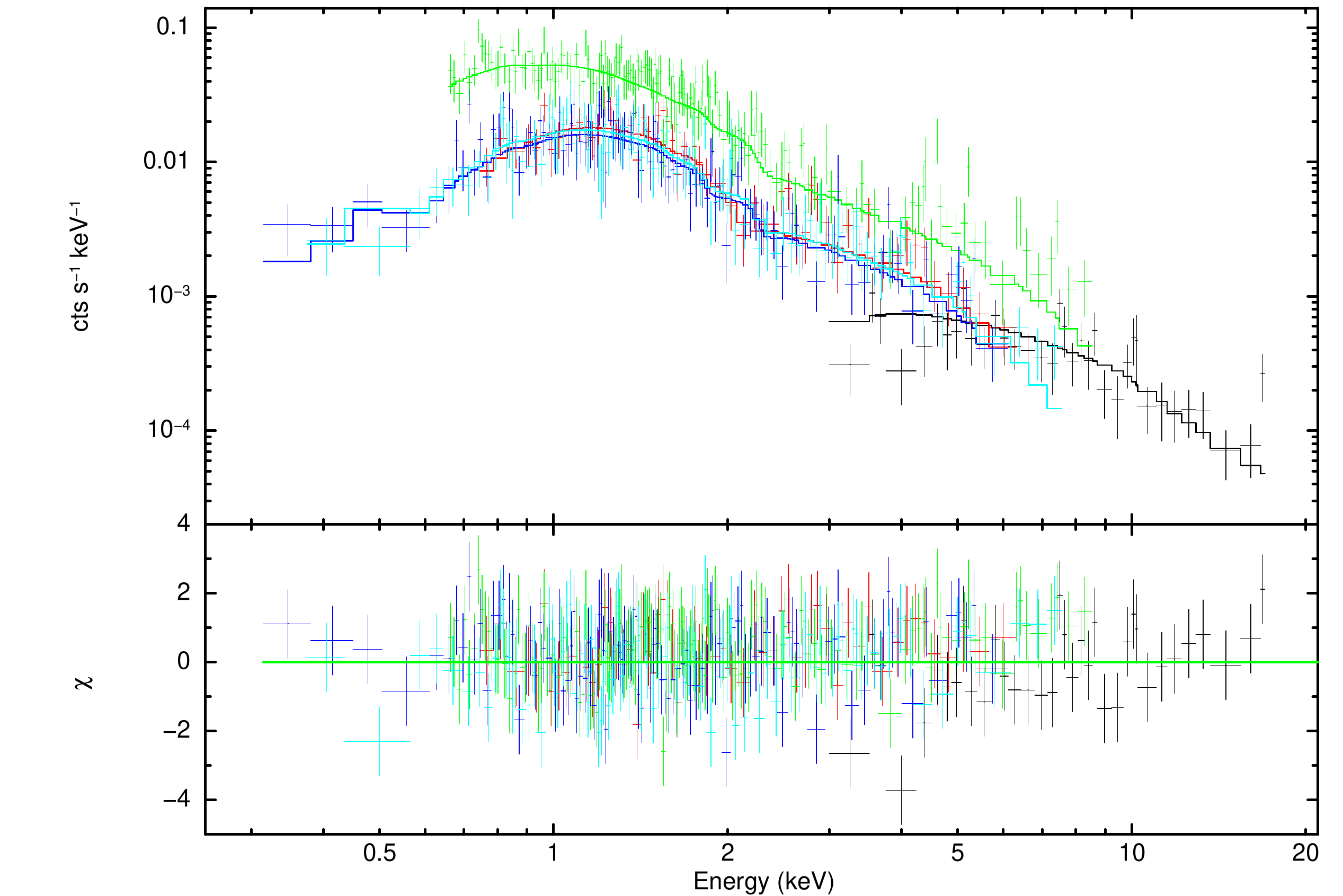}%
    \caption{\label{fig:jointfit} \chandra, \xmm, and \nustar\ spectra jointly fit with an absorbed powerlaw plus blackbody model. Black: \nustar; Red: \chandra; Green, Turquoise, and Purple: \xmm\ EPIC PN, MOS 1, and MOS 2, respectively.}%
    \label{fig:xrayspectra}%
\end{figure}

\begin{deluxetable*}{c c c c c c c c}
\centering
\tablecaption{X-ray joint fit spectral parameters. All errors are given with $1\sigma$ confidence. \label{tab:jointfit}}
\tablehead{
\colhead{Model} & \colhead{$N_H$, $10^{21}$ cm\textsuperscript{-2}} & \colhead{T, eV} & \colhead{R, km} & \colhead{$\Gamma$} & \colhead{powerlaw norm, $10^{-5}$} & \colhead{$\chi^2_v$} & \colhead{dof} 
}
\startdata
BB & $3.3_{-0.8}^{+1.2}$ & $180_{-40}^{+50}$ & $0.5^{-0.3}_{+0.8}$ & $2.0 \pm 0.1$ & $6.7_{-1.3}^{+1.4}$ & 1.0 & 482 \\
NSMAX ($B=10^{12}$ G) & $4.0^{+0.2}_{-0.8}$ & $28^{+6}$ & $10^{+3}_{-9}$ & $1.9 \pm 0.1$ & $5.8^{+0.5}_{-0.7}$ & 1.0 & 482 \\
\enddata
\end{deluxetable*}

When fitting, the \nustar\ spectrum was cut off above 20 keV where the background began to dominate. Thus we used a range of 3 to 20 keV for the \nustar\ spectrum, a range of 0.5 to 10 keV for \chandra\, and a range of 0.2 to 10 keV for \xmm\ EPIC PN, MOS 1, and MOS 2. All error bars were calculated to $1\sigma$ significance. We fit the \nustar, \chandra\, and \xmm\ data jointly with an absorbed power-law model plus a blackbody component \texttt{(tbabs*(bbodyrad+powerlaw))} to account for the central pulsar \citep[abundances from][]{Wilms2000}. This fit (Fig. \ref{fig:jointfit}) produced a photon index of $\Gamma = 2.0 \pm 0.1$ (typical for a PWN), a neutral hydrogen absorption of $N_H = 3.3_{-0.8}^{+1.2} \times 10^{21}$ cm\textsuperscript{-2}, a blackbody temperature of $T = 0.18_{-0.04}^{+0.05}$ keV, and a blackbody radius of $R = 0.5_{-0.3}^{+0.8}$ km (Table \ref{tab:jointfit}). All results are consistent with \cite{Karpova}, who theorized that this blackbody radius and temperature model emission from a hot polar cap. Absorbed X-ray flux was $2.4 \pm 0.1 \times 10^{-13}$ erg s\textsuperscript{-1} cm\textsuperscript{-2} in the 2 to 20 keV band and $2.4^{+0}_{-0.2} \times 10^{-13}$ erg s\textsuperscript{-1} cm\textsuperscript{-2} in the 0.5 to 8 keV band, with $L_{2-20} = 2.9_{-0.3}^{+0.2} \times 10^{31}$ erg s\textsuperscript{-1} (unabsorbed) at a distance of 1 kpc. This fit resulted in a reduced chi-squared of 1.0 for 482 degrees of freedom, confirming that the spectrum is non-thermal and fits well to a single power-law model. There is no evidence of a spectral cutoff or break up to 20 keV. 

The spectra were also fit with a neutron star atmosphere model as a replacement for the blackbody model \texttt{(tbabs*(NSMAX+powerlaw))} with $B = 10^{12}$ G to model thermal emission from the entire neutron star surface. This fit resulted in a photon index of $\Gamma = 1.9 \pm 0.1$, a neutral hydrogen absorption of $N_H = 4.0^{+0.2}_{-0.8} \times 10^{21}$ cm\textsuperscript{-2}, an atmosphere temperature of $T = 28^{+6}$ eV, and a radius of $R = 10^{+3}_{-9}$ km (Table \ref{tab:jointfit}). No lower limit was derived for the temperature, as the temperature fit to the lower limit value of the model. The gravitational redshift was frozen to a standard value of 0.3. The NSMAX fit is not shown in Fig. \ref{fig:xrayspectra}, as it resulted in similar residuals and goodness of fit as the blackbody model. 

For SED fitting, an annulus of $r = 2$" centered on the central point source was subtracted from the \chandra\ data to avoid thermal emission from the putative neutron star, and photons below 2 keV (where the absorption component cut off) were omitted to account for neutral hydrogen absorption at low energies. Both the NS thermal emission and ISM absorption are negligible in the \nustar\ band (3-79 keV). Thus, the X-ray data (\chandra\ and \nustar\ only) fit in the SEDs contains only the power-law contribution.

\subsubsection{Spatially Resolved Spectral Analysis with Chandra}

\begin{figure}%
    \centering
{{\includegraphics[width=0.9\columnwidth]{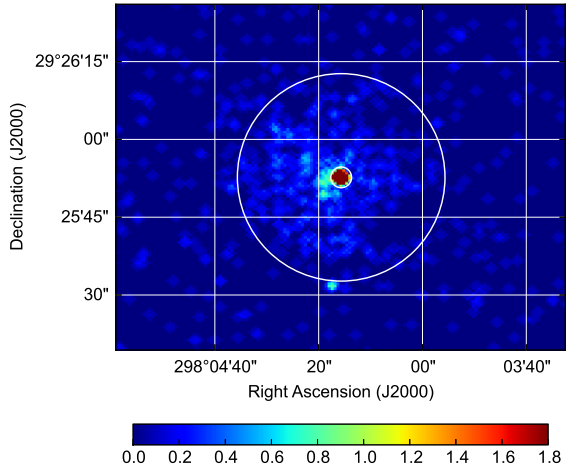} }}%
    \caption{\chandra\ image of the DA~495 wind nebula, 0.5 - 8 keV, smoothed with a 
    Gaussian kernel to $3\sigma$ significance. Inner white 2" circle denotes pulsar location, outer 40" white circle denotes wind nebula extent.}%
    \label{fig:chandra}%
\end{figure}

\chandra's high angular resolution (\textapprox 0.5") was used to investigate potential spectral softening at larger distances from the central pulsar due to synchrotron burnoff in the pulsar's leptonic wind. Spectra were extracted from two annuli: 2" $\rm < r_1 <$ 10" and 10" $\rm < r_2 <$ 20". The inner 2" of the nebula were ignored to avoid the contribution of blackbody emission from the putative central pulsar. Each annulus spectrum was extracted using \texttt{CIAO 4.10} extended emission procedures and was fit in \texttt{XSPEC} from 0.5 to 8 keV with an absorbed power law model \citep[again, using abundances from][]{Wilms2000}. The fitting resulted in an $N_H$ of $4.8_{-2.0}^{+2.6} \times 10^{21}$ cm\textsuperscript{-2} and a photon index of $1.9 \pm 0.2$ for $\rm r_1$ ($\chi_{v}^{2} = 0.6, 13$ dof), while $\rm r_2$ gave an $N_H$ of $4.8_{-2.4}^{+3.7} \times 10^{21}$ cm\textsuperscript{-2} and a photon index of $1.9 \pm 0.3$ ($\chi_{v}^{2} = 0.9, 15$ dof). These $N_H$ and $\Gamma$ values are consistent with the full \chandra\ spectrum extracted from the region 2" $\rm < r <$ 20". To better constrain the photon indices, we refit the two annuli with $N_H$ frozen to $3.3 \times 10^{21}$ cm\textsuperscript{-2}, the full X-ray spectrum joint fit value using thermal component \texttt{bbodyrad} (Table \ref{tab:jointfit}). This resulted in a photon index of $\Gamma = 1.7^{+0.1}_{-0.1}$ for both annuli. We did not find any variation in $\Gamma$ between the two annuli.

A radial profile of the \chandra\ data was generated using an inner radius of 2", an outer radius of 20", and 5 annuli. We plotted the brightness of the wind nebula as a function of distance from the central point source in the energy bands 0.5-1.0 keV, 1.0-3.0 keV, and 3.0-8.0 keV, normalizing the data points at r = 2" to 1.0 and subtracting the background. We saw no significant spectral hardening or softening between the three energy bands. In the annular fits, the photon index stays constant within error bars at larger distances from the central pulsar, consistent with the radial profile results. We report no evidence of quickening in the burnout of electrons at higher energies. 



\subsubsection{X-ray Flux Upper Limit}

\begin{figure}%
    \centering
    \subfloat[]{\includegraphics[width=0.45\textwidth]{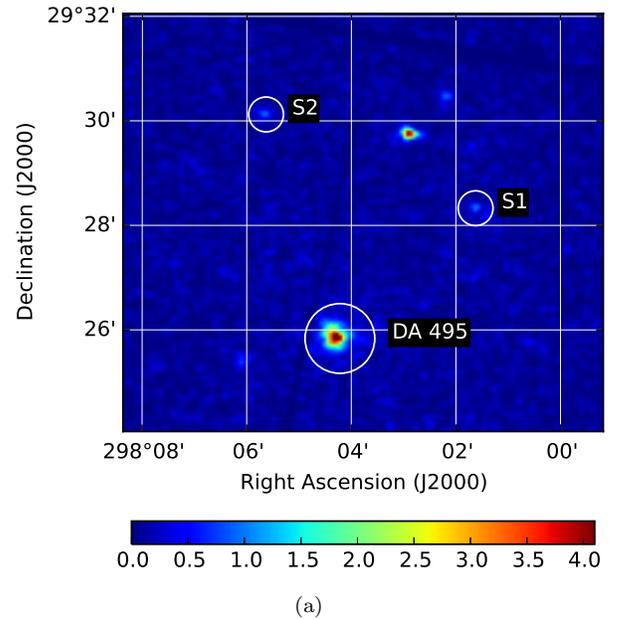}}%
    \qquad
    \subfloat[]{{\includegraphics[width=0.45\textwidth]{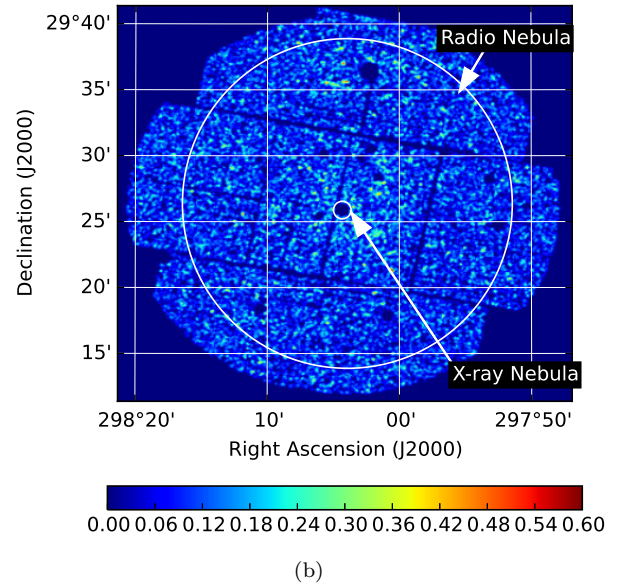}}}%
    \caption{(a) XMM-MOS 2 image of DA 495, 0.5-10 keV. (b) XMM-MOS 2 background subtracted and point-source masked image, 2-10 keV. Both images were smoothed with a 
    Gaussian kernel to $3\sigma$ significance.}%
    \label{fig:xmm}%
\end{figure}

An X-ray flux upper limit for SED fitting was extracted by analyzing the extended emission within the radio nebula region r \textless\ 12.5' (Fig. \ref{fig:xmm}b). In a single electron population leptonic emission scenario one would expect diffuse X-ray emission coincident with the large radio nebula. \xmm\ was the only X-ray telescope with a large enough field of view to be used to attempt a diffuse nebula upper limit. EPIC-MOS 2 was the sole camera used for analysis, as two CCD chips in MOS 1 were compromised, and EPIC-PN was operated in the Small Window mode. \texttt{XMM SAS 17.0.0 ESAS} procedures were used for source and background spectrum generation. ESAS background subtraction procedures account for the quiescent particle background (QPB), but do not account for CXB emission, solar wind charge exchange (SWCX) background, soft proton background, or instrumental lines. Because instrumental lines and SWCX effects manifest below 2 keV, all energies below 2 keV were ignored. The lightcurve of the region was examined and no flares were found. The CXB was accounted for by fitting a second power-law to the spectrum using the known photon index of $\Gamma = 1.46$ \citep{Snowden2004}. It was not possible to directly subtract a local background spectrum, as the radio nebula covers the entire XMM field of view and therefore no region was guaranteed to be source-free. All point-like sources, including DA~495, were masked using the \texttt{XMM ESAS cheese} command, and the extended emission spectrum was fit separately from the compact nebula. Because of the sharp (factor of \textapprox 6) drop in flux between the compact X-ray nebula and surrounding background (Fig. \ref{fig:xmm}), it is unlikely that much of the measured X-ray flux from the radio emission region is from DA~495.

When fitting the extended region, $N_H$ was frozen to the joint fit value of $3.3 \times 10^{21}$ cm\textsuperscript{-2}. Subtracting the relevant background components and ignoring the photons below 2 keV resulted in a flux (2-10 keV) of $3.6^{+0.7}_{-0.5} \times 10^{-12}$ erg s\textsuperscript{-1} cm\textsuperscript{-2} when the DA~495 photon index was allowed to fit freely (very soft at $\Gamma \approx 5.6$), and a flux of $\approx 4.7 \times 10^{-12}$ erg s\textsuperscript{-1} cm\textsuperscript{-2} when the photon index was frozen to the DA~495 best fit value of 2.0 (with the normalization of the CXB component frozen to the value found in the previous fit). The compact X-ray nebula upper limit flux was measured to be $\approx 1.6 \times 10^{-13}$ erg s\textsuperscript{-1} cm\textsuperscript{-2} in the 2-10 keV band. The extended region flux (with all point-like sources masked) was added to the DA~495 compact flux value to gain a diffuse X-ray nebula upper limit of $(3.3 - 4.9)\times 10^{-12}$ erg s\textsuperscript{-1} cm\textsuperscript{-2}. 

\section{DA~495 PWN Spectral Modeling} \label{sec:model}

\begin{figure}[hbt]
    \centering
    \vspace{-5pt}
    \includegraphics[width=\linewidth]{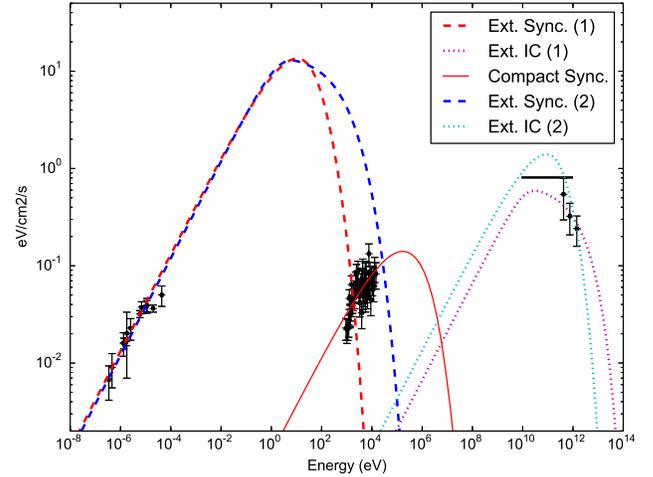}
    \vspace{-5pt}
    \caption{Leptonic model 1 and 2. The extended radio and TeV nebulae are fit with inverse Compton scattering by primary electrons (magneta dot for model 1 and cyan dot for model 2) and primary electron synchrotron (red dash for model 1 and blue dash for model 2). The compact X-ray nebula is fit with another primary electron synchrotron (red solid). Parameters are listed in Table \ref{table}. Black points are the radio \citet{2008ApJ...687..516K}, X-ray (this work) and VERITAS data points \citep{2018ApJ...866...24A}. Black solid line is the {\it Fermi} upper limits \citep{2018ApJ...866...24A}.}
    \label{fig:lepto1}
\end{figure}

\begin{figure}[hbt]
    \centering
    \vspace{-5pt}
    \includegraphics[width=\linewidth]{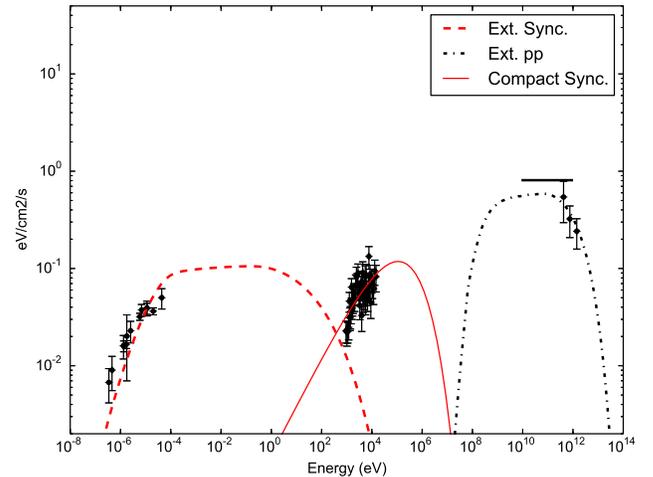}
    \vspace{-5pt}
    \caption{Hadronic fit of the broadband SED. The extended radio and TeV nebulae are fit with pp collisions of primary protons (black dash dot) and secondary pair synchrotron (red dash). The compact X-ray nebula is fit with the primary electron synchrotron (red solid). Parameters are listed in Table \ref{table}. Black points are the radio \citet{2008ApJ...687..516K}, X-ray (this work) and VERITAS data points \citep{2018ApJ...866...24A}. Black solid line is the {\it Fermi} upper limits \citep{2018ApJ...866...24A}.}
    \label{fig:hadronic}
\end{figure}

The multi-wavelength spectra and spatial extents of DA~495 PWN provide strong constraints on the underlying radiation processes. In particular, since the X-ray nebula has significantly smaller extent than the radio and TeV nebulae, it naturally argues for a two-zone model, where the inner X-ray nebula originates from recent acceleration of primary electrons in the neighborhood of the central pulsar with a probably higher magnetic field, while the more extended radio and TeV nebulae are the consequence of older particles that have diffused away from the central pulsar. Given the very high flux of the extended TeV nebula, there could be a hadronic contribution to the high-energy spectral component. In this section, we discuss the fitting models for the broadband spectrum of the extended radio and TeV nebulae as well as the compact X-ray nebula. We use both the spectrum and spatial extent to constrain the model parameters. Fitting results are shown in Figures \ref{fig:hadronic} and \ref{fig:lepto1} with model parameters listed in Table \ref{table}.

\begin{table*}[ht]
    \centering
    \begin{tabular}{|c|c|c|c|} \hline
Extended radio and TeV nebulae parameters & Leptonic Model 1 & Leptonic Model 2 & Hadronic Model \\ \hline
Distance                        & \multicolumn{3}{c|}{1 kpc}                                     \\ \hline
Magnetic field                  & 10 $\rm{\mu G}$          & 15 $\rm{\mu G}$    & 0.8 mG         \\ \hline
Electron minimal Lorentz factor & \multicolumn{2}{c|}{1}                        & --             \\ \hline
Electron maximal Lorentz factor & $10^7$                   & $5\times 10^7$     & --             \\ \hline
Electron power-law index        & 2.0                      & 2.0                & --             \\ \hline
Electron total kinetic energy   & $2\times 10^{47}$ erg    & $10^{47}$ erg      & --             \\ \hline
Proton minimal Lorentz factor   & --                       & --                 & 1.0            \\ \hline
Proton maximal Lorentz factor   & --                       & --                 & $2\times 10^4$ \\ \hline
Proton power-law index          & --                       & --                 & 2.0            \\ \hline
Proton total kinetic energy     & --                       & --         & $7\times 10^{48}$ erg  \\ \hline
Target proton density           & --                       & --         & $1~\rm{cm^{-3}}$       \\ \hline
Diffusion coefficient           & \multicolumn{2}{c|}{$4\times 10^{25}~\rm{cm^2s^{-1}}$} & $1.5\times 10^{25}~\rm{cm^2\,s^{-1}}$ \\ \hline \hline
Compact X-ray nebula parameters & \multicolumn{2}{c|}{ Leptonic Model}           &  Hadronic Model \\ \hline
Magnetic field                  & \multicolumn{2}{c|}{$50~\rm{\mu G}$}  & 0.8 mG                 \\ \hline
Electron minimal Lorentz factor & \multicolumn{3}{c|}{1}                                         \\ \hline
Electron maximal Lorentz factor & \multicolumn{2}{c|}{$5\times 10^8$}           & $10^8$         \\ \hline
Electron power-law index        & \multicolumn{3}{c|}{2.0}                                       \\ \hline
Electron total kinetic energy   & \multicolumn{2}{c|}{$2\times 10^{42}$ erg} & $3\times 10^{40}$ erg \\ \hline
    \end{tabular}
    \caption{Fitting parameters of leptonic and hadronic models. Notice that our radiation code considers radiation cooling, which may introduce a cooling break in the power-law spectrum.}
    \label{table}
\end{table*}

\subsection{Pure Leptonic model} \label{sec:purelepton}

In a pure leptonic model, the radio to TeV emission generally consists of two components, namely, a low-energy component from synchrotron emission from primary electrons and a high-energy component from inverse Compton scattering of CMB photons by the same electrons. Given the large extents of the radio and TeV nebulae, the  primary electron synchrotron photon density is much lower than the CMB. Thus we do not expect a significant synchrotron-self Compton contribution. The compact X-ray nebula originates from the neighborhood of the central pulsar, where the higher magnetic field and freshly accelerated electrons trigger an additional synchrotron component.

We can estimate the average magnetic field strength within the radio and TeV nebulae based on the spectral shapes. In a pure leptonic model, the radio and TeV emission should come from the same electron population. Since the radio spectrum is rising in a power-law shape while the TeV spectrum appears like a cutoff, we expect that the underlying electron spectrum has a spectral break at some maximal energy. To upscatter CMB photons to TeV energies, the primary electron cutoff should be $\sim 20~\rm{TeV}$. Generally speaking, this spectral break can have two origins, either the synchrotron cooling break or the intrinsic spectral cutoff. If the electron cutoff at $\sim 20~\rm{TeV}$ results from synchrotron cooling, the synchrotron cooling time scale should be comparable to the source age,
\begin{equation}
\begin{aligned}	
     t_{cool} & = \frac{m_ec^2}{4/3c\sigma_T\gamma (u_B+u_{CMB})} \\ 
     		   & = 0.976~\rm{yr}~ \gamma^{-1}(u_B+u_{CMB})^{-1}\sim t_{age}=20~\rm{kyr}~~.
\end{aligned}
     \label{equation:syncool}
\end{equation}
Therefore, we can find that $B \sim 8~\rm{\mu G}$, consistent with the ISM magnetic field. This implies that the synchrotron component should peak at
\begin{equation}
     h\nu_c=h\frac{3eB}{4\pi m_e c}\gamma^2\sim1.74\times 10^{-8}~\rm{eV}~B\gamma^2 \sim 50~\rm{eV}~~,
     \label{equation:syncrit}
\end{equation}
which is in the ultraviolet.

The extent of the TeV nebula is likely determined by the diffusion of TeV electrons. The typical interstellar diffusion coefficient of $\sim 10~\rm{TeV}$ electrons is $\sim 10^{29}~\rm{cm^2s^{-1}}$. Considering the radiative cooling of these TeV electrons in the ISM, the diffusion radius is given by
\begin{equation}
     d_{dif}=2\sqrt{Dt_{cool}}\sim 100~\rm{pc}.
\end{equation}
Apparently, the extent should be much larger than the observed angular extent assuming a distance of $1~\rm{kpc}$. To trap the high-energy electrons within the observed $\sim 2~\rm{pc}$ region, the diffusion coefficient should be $D\sim 4\times 10^{25}~\rm{cm^2s^{-1}}$, much smaller than the ISM diffusion coefficient. Comparing to the Bohm diffusion coefficient, which is given by
\begin{equation}
    D_{Bohm}=\frac{1}{3}\frac{\gamma mc^2}{eB}c \sim 6.67\times 10^{25}~\rm{cm^2\,s^{-1}}~(\frac{B}{5~\rm{\mu G}})^{-1}~~,
\end{equation}
the observed slow diffusion then sets a lower limit on the average magnetic field, $B\gtrsim 10~\rm{\mu G}$. Recent HAWC observation of the Geminga PWN suggested a similarly small diffusion coefficient within the Geminga nebular diffusion comparable to the local Bohm diffusion coefficient \citep{2017Sci...358..911A,2018APh...102....1L}, which is consistent with our DA~495 fitting parameters. 

Since the same electrons produce the radio emission through synchrotron and the TeV emission through inverse Compton scattering, their flux levels can be used to derive the ratio between the magnetic energy density and the target photon energy density, namely, $u_B/u_{CMB}$. Figure \ref{fig:lepto1} shows our leptonic fitting results. We find that the magnetic field cannot be too high, or else there should exist a very bright X-ray nebula of similar size to the radio and TeV nebulae, which cannot be confirmed in our observations. Notice that our spectral fitting considers cooling effects, thus we observe a cooling break in the $B=15~\rm{\mu G}$ case.

The sharp cutoff at the edge of the central compact X-ray nebula implies a stronger magnetic field and newly accelerated electrons. Figure \ref{fig:lepto1} shows a sample fitting with $50~\rm{\mu G}$ magnetic field, although with merely X-ray data we cannot constrain both the magnetic field strength and the electron spectrum. The higher magnetic field strength should infer brighter radio emission within the compact X-ray nebula. Instead, K08 suggests an apparent radio hole in that region. This casts doubt on simple leptonic models.

\subsection{Pure Hadronic Model}\label{sec:purehadron}

In a pure hadronic model, the broadband emission of the extended nebula originates from proton-proton (pp) collisions, where the neutral pion decay makes the high-energy spectral component while the charged pion decay results in secondary electron-positron pairs, which then give rise to the low-energy spectral component through synchrotron radiation. The compact X-ray nebula is likely the primary electron synchrotron radiation, and its small extent is due to the fast cooling time of these electrons in a high magnetic field.

The observed radio and TeV spectra can put strong constraints on the underlying proton spectral distribution and magnetic field strength. The soft TeV spectral shape indicates the neutral pion decay spectral cutoff. This corresponds to a proton cutoff energy at \citep{2006PhRvD..74c4018K}
\begin{equation}
E_{p,cut}=10\times E_{\gamma,cut}\sim 10~\rm{TeV}~~.
\end{equation}
Meanwhile, the rising radio spectrum marks the low-energy cutoff of the secondary electrons. These electrons, which come from the charged pion decay, should have a low-energy cutoff at $\sim 500~\rm{MeV}$ \citep{2006PhRvD..74c4018K}. This is because the cross section of the pp collisions cuts off when the nonthermal protons become nonrelativistic. Therefore, we can quickly derive from Equation \ref{equation:syncrit} that the magnetic field strength should be
\begin{equation}
B\sim 0.5~\rm{mG}~\frac{E_{radio}}{10^{-5}~\rm{eV}}(\frac{\gamma}{10^3})^{-2}.
\end{equation}
Since the cross section for charged pions is about half of the neutral pion cross-section, and part of the charged pion energy goes into neutrinos, the total electron power from pp collisions is typically about one third of the total $\gamma$-ray power. We can see from the DA~495 spectrum that the TeV flux is $\sim$5 times higher than the radio flux, indicating that the secondary synchrotron emission is very efficient (Figure \ref{fig:hadronic}). Indeed, we can quickly estimate from Equation \ref{equation:syncool} that the cooling time of the secondary electrons in a $\sim 0.5~\rm{mG}$ magnetic field should be $\sim 90~\rm{kyr}$ for the low-energy cutoff electrons at $\sim 500~\rm{MeV}$, on the same order of the source age.

Given the extent of the TeV nebula, we can estimate the diffusion coefficient inside. Since the protons that produce the TeV emission do not cool in the nebula, the diffusion coefficient is
\begin{equation}
D=(\frac{R}{2})^2/t_{age}\sim 1.5\times 10^{25}~\rm{cm^2\,s^{-1}}~~,
\end{equation}
a little smaller than that in the leptonic scenario. But here the magnetic field is two orders of magnitude higher than the typical ISM value, thus the diffusion coefficient is well above the Bohm limit.

While the radio and TeV nebulae are produced by pp collisions, the compact X-ray nebula near the central pulsar comes from the primary electrons co-accelerated with the protons. Due to the high magnetic field, these electrons will cool in a short time, which naturally explains the small spatial extent of the X-ray nebula. From Equation \ref{equation:syncrit}, the X-ray spectral shape implies that the primary electrons should be accelerated to $\sim 20~\rm{TeV}$, consistent with the maximal proton energy.

\section{Discussion} \label{sec:discussion}

With the detection of high-energy X-ray and TeV emission, we have modeled several different scenarios that explain the nature of the broadband emission from DA~495. K08 and \citet{2008ApJ...687..505A} describe DA~495 as a PWN possibly unconstrained by the presence of a reverberation supernova reverse shock, allowing for the $\sim$20 kyr PWN to expand with an $R_{radio}/R_{X-ray}\sim 10\times$ that of PWN of similar age ($\sim 20\rm\ kyr$). It is also understood that the radio region is highly magnetized with a strong dipole field and that the dipole axis aligns with asymmetries in the X-ray observations likely due to a jet from the pulsar. 

Even before the hard X-ray and TeV detection, DA~495 was an enigmatic PWN, and our modeling with the newest observations continues to support this idea - as well as opens up some new questions. There now exists tension between the current understanding and the SED modeling scenarios that we presented in the previous section, particularly once we include the TeV emission. In this section we will discuss some of these conflicts and explain future studies that could shed light on the emission processes.

\subsection{Estimated Age}
Certainly, without a detected pulsar the age estimation becomes more challenging and the displacement of the neutron star from the radio nebula's center, presuming a reasonable kick, is consistent with the currently estimated ages \citep{Arzoumanian2004}. There are several relationships for PWN that can place rough age constraints on the undetected pulsar based on observational parameters alone. 

A study on the X-ray and $\gamma$-ray luminosity of PWN \citep{2009ApJ...694...12M} describes a relationship between $L_{1-30\ TeV} / L_{2-10\ keV}$ that can be used to find $\tau_c$ of the pulsar and $\dot{E}$. This is a characteristic relationship due to the different electron populations that generate the X-ray synchrotron and TeV inverse-Compton emissions. Using the 1~kpc luminosity from the VERITAS spectral fit and the joint X-ray fit in this work: $L_{1-30\ TeV} = (1.3\pm 0.2) \times 10^{31} \rm\  erg\ s^{-1}$ and $L_{2-10\ keV}=(2.6 \pm 0.2) \times 10^{31} \rm\  erg\ s^{-1}$ for a ratio of $L_{1-30\ TeV} / L_{2-10\ keV} = 0.50 \pm 0.05$. This results in a derived $\dot{E} = (2.6\pm 0.2) \times 10^{37} \rm\ erg\ s^{-1}$ and $\tau_c = 3.5 \pm 0.5$ kyr -- dramatically different than the K08 estimated values of $\dot{E} \sim 10^{35} \rm\ erg\ s^{-1}$ and $\tau_c \sim 20$ kyr. The authors in \citet{2009ApJ...694...12M} note that DA~495 is expected to be an exception with respect to their fits due to the strong magnetic field inferred from radio observations. Now that we have included the TeV emission, we can confirm their suspicions that DA~495 does not line up with the rest of the PWN population from an observational perspective. If the younger age estimate is to be believed, then we would expect a shell of angular size $\theta_{SNR} \approx 31\arcmin$ (using Equation (1) from \citet{2005ApJ...631..480R}). Without clear evidence for a radio shell at this size, we do not find this estimated age to be more reliable than the synchrotron cooling age that was estimated by K08.

Another population study of PWNe was performed by HESS \citep{2018A&A...612A...2H}, where we get yet another story about the age. For this study, TeV luminosity ($L_{1-10\ TeV}$) was used to find estimate the age, spindown luminosity, and size. Using the K08 age for DA~495, the expected $L_{1-10\ TeV}\rm\ (@20\ kyr\ and\ 1\ kpc) \rm\ is \sim 7\times 10^{33} \rm\ erg\ s^{-1}$. The measured $L_{1-10\ TeV}\rm\ (@1\ kpc)\ =(1.3\pm 0.2) \times 10^{31} \rm\  erg\ s^{-1}$ would put the age at $>$1 Myr when using the HESS flux. So, while the enhanced TeV emission is causing issues with understanding the SED models with a strong magnetic field, by these estimates the TeV emission is very low for the expected age as the TeV luminosity is expected to decrease with the PWN age. There is some suggestion in K08 that DA~495 is at a larger distance ($D\sim 5$ kpc). The measured luminosity,  $L_{1-10\ TeV}\rm\ (@1\ kpc)\ =(3.3\pm 0.4) \times 10^{31} \rm\  erg\ s^{-1}$,  still has significant disagreement with any assumed age. Without any other age estimates, the luminosity at 5 kpc would still put the age of DA~495 at $>$1 Myr. We are not suggesting that DA~495 is a remnant that old, but that it is an outlier from an observational standpoint and that standard PWN evolutionary scenarios do not apply. 

These conflicting results on the age of the object reinforce our conclusions that DA~495 has not followed a typical evolutionary path. Time-dependent modeling of atypical evolutionary scenarios of the PWN including SNR evolution would be the most likely way to help resolve some of the questions regarding possible age.

\subsection{PWN or Shell-SNR?}
\label{pwnorsnr}

The magnetic field strength estimated from the radio observations and the SED may be reconciled using a model including hadronic emission. This requires hadronic target material and the presence of a SNR shell. This is contrary to the K08 claim about the lack of a shell, though earlier interpretation of the radio emission from \citet{1989JApA...10..161V} argued that DA~495 is a composite remnant with a thick shell due to slow supernova ejecta interacting with the reverse shock. The thick shell claim may still be considered a possibility if there is a superimposed toroidal magnetic field as K08 suggests. The magnetic field strength of the hadronic scenario may be possible with there only being a shock interacting with ejecta in a ring around the source. This, in part, could also explain the very low surface brightness of the nebula, since the radio and TeV emission would be coming only from the equatorial ring. However, the physical reason how the toroidal magnetic field was entrapped is not clear. This interpretation still has issues within the context of other observations as there is not an obvious, cooler, dense, target region as expected with most shock interactions. 


A way to further investigate the existence of a shell-type remnant would be to examine the TeV morphology more closely. The radio hole is about 2$\arcmin$ in diameter, and if the radio and TeV emission both share this feature, then this could indicate that the emission is from a shocked region where the magnetic field was compressed. The putative hadronic targets in this possible scenario would be slower ejecta from the supernova.
If the TeV emission is smooth through the middle of the region, then that would indicate that the electrons from the central pulsar are likely responsible for the TeV emission. This argues against the hadronic scenario, since the radio and TeV emissions are no longer spatially linked. Currently, none of the gamma-ray instruments are sensitive enough to definitively make this distinction, but the Cherenkov Telescope Array (CTA), due to come online in the next few years, may have the enhanced angular resolution necessary \citep{2017arXiv170907997C}.

\subsection{Radiation Mechanism and Magnetic Field}

While both leptonic and hadronic scenarios can reasonably reproduce the multi-wavelength observation (Figures \ref{fig:hadronic} and \ref{fig:lepto1}), we can quickly notice several key predictions from the spectral modeling. For a pure leptonic model, we expect a straight power-law in the low-energy spectral component, from radio to optical. K08, however, suggested a spectral break in the radio component. Future observations at higher radio frequencies to far infrared may help to diagnose the significance of this spectral break. Additionally, the pure leptonic model predicts rather high optical to ultraviolet flux. Most interestingly, it implies that there exists a diffusive X-ray nebula that is of similar size as the radio and TeV nebulae. The total flux is comparable to the central compact X-ray nebula, but since it spreads out $\sim 2~\rm{pc}$, its surface brightness can be much lower than the central compact X-ray nebula, which is very hard to detect. The leptonic model requires an average magnetic field that is slightly higher than the typical ISM value, probably due to the interaction between pulsar wind and the ISM, or maybe the remnant magnetic field from the supernova. The total energy budget for the leptonic model is $\sim 10^{48}~\rm{erg}$, typical for PWNe.

The hadronic models lead to very different predictions. In a pure hadronic model, the radio spectral break is a natural result of the secondary pair synchrotron. Both the infrared to optical and {\it Fermi} $\gamma$-ray spectra should appear flat, assuming an underlying proton energy spectral index of $\sim 2.0$. The pure hadronic model also suggests that there is no diffusive X-ray nebula, which is consistent with our observations. However, the hadronic model requires a very strong magnetic field pervading the radio and TeV nebulae, which is likely powered by the central pulsar, consistent with K08. Nonetheless, protons are generally unlikely to be accelerated at typical particle acceleration sites near the pulsar, such as the polar cap. Instead, they should be accelerated due to the interaction between the pulsar wind and the ISM, or magnetic reconnection in the highly magnetized nebula. Given the fitting parameters in Table \ref{table}, we can see that the total magnetic energy within the PWN is about $2.5\times 10^{49}~\rm{erg}$, indicating that the nebula is considerably magnetized. These features require that DA~495 is a very unusual PWN, one which has extremely high power and can extend its magnetic field to $\sim 2~\rm{pc}$ into the ISM. Interestingly, the latter is supported by the radio polarization map by K08, where they find a dipole shape magnetic field morphology in the radio nebula.




\section{Conclusion}

In this paper we presented new observations from \nustar . We combined this new analysis with recent TeV gamma-ray observations and the current radio analysis to create a broadband spectral energy distribution for DA~495. Using analytical modeling we described several scenarios for the particle population within the PWN, and put the modeling in context with previous discussion about DA~495's nature described in (\textit{eg}) \citet{2008ApJ...687..516K}.

DA~495 is a unique PWN, still with many unknowns. We find some evidence for a non-PWN scenario that could interpret the radio and TeV emission as a thick shell containing relativistic hadrons -- possibly accelerated from the supernova shock that is interacting with some slow supernova ejecta. This challenges the current interpretation and requires further investigation by future TeV gamma-ray observations. With better angular sensitivity, comparisons between radio and TeV morphology could provide evidence regarding this conclusion.

\section*{Acknowledgements}
The authors would like to thank S. Reynolds and M. Pohl for their insightful discussions. This work used data from the \nustar\ mission, a project led by the California Institute of Technology, managed by the Jet Propulsion Laboratory, and funded by NASA. We made use of the \nustar\  Data Analysis Software (NuSTARDAS) jointly developed by the ASI Science Data Center (ASDC, Italy) and the California Institute of Technology (USA). HZ acknowledges the support from Fermi Guest Investigator program, Cycle 11, grant number 80NSSC18K1723. TBH acknowledges the generous support of the National Science Foundation under cooperative agreement PHY-1352567. HL acknowledges the support by the LANL/LDRD program.

\end{document}